\documentclass[journal, twoside, letterpaper, twocolumn]{IEEEtran}
\pdfoutput=1

\usepackage{cite}

\ifCLASSINFOpdf
  \usepackage[pdftex]{graphicx}
  \graphicspath{{./}}
\else
\fi

\usepackage{amsmath}

\usepackage{url}

\usepackage{epstopdf}
\usepackage{bm}

\newcommand{\mat}[1]{\mathbf{#1}}
\newcommand{\matc}[2]{\begin{pmatrix}{#1}\\{#2}\end{pmatrix}}

\newcommand{\matd}[4]{\begin{pmatrix}{#1}&{#2}\\{#3}&{#4}\end{pmatrix}}

\usepackage[normalem]{ulem, xcolor}

\usepackage{array}
\newcolumntype{P}[1]{>{\centering\arraybackslash}p{#1}}
\newcolumntype{M}[1]{>{\centering\arraybackslash}m{#1}}

\newcommand{\addportrait}[1]{\includegraphics[width=1in,height=1.25in,clip,keepaspectratio]{#1}}

\begin{document}
%
\title{Numerical Investigation of Vertical Cavity Lasers with Subwavelength Gratings Using the Fourier Modal Method}

\author{Alireza~Taghizadeh,~\IEEEmembership{Member,~IEEE,} Jesper M{\o}rk, and~Il-Sug~Chung*~
\thanks{Manuscript received XXX XX, XXXX; revised XXX XX, XXXX; accepted XXX XX, XXXX. Date of current version XXX XX, XXXX.}
\thanks{A. Taghizadeh, J. M{\o}rk and I.-S. Chung are with the Department of Photonics Engineering (DTU Fotonik), Technical University of Denmark, DK-2800 Kgs. Lyngby, Denmark. *e-mail: ilch@fotonik.dtu.dk.}}

\markboth{\uppercase{Journal of Lightwave Technology}, ~Vol.~XX, No.~X, XXX~XXXX}%
{Taghizadeh \MakeLowercase{\textit{et al.}}: \uppercase{Numerical investigation of vertical cavity lasers with subwavelength gratings}}
\maketitle

\begin{abstract}
We show the strength of the Fourier modal method (FMM) for numerically investigating the optical properties of vertical cavities including subwavelength gratings. Three different techniques for determining the resonance frequency and Q-factor of a cavity mode are compared. Based on that, the Fabry-Perot approach has been chosen due to its numerical efficiency. The computational uncertainty in determining the resonance frequency and Q-factor is investigated, showing that the uncertainty in the Q-factor calculation can be a few  orders of magnitude larger than that in the resonance frequency calculation. Moreover, a method for reducing 3D simulations to lower-dimensional simulations is suggested, and is shown to enable approximate and fast simulations of certain device parameters. Numerical calculation of the cavity dispersion, which is an important characteristic of vertical cavities, is illustrated. By employing the implemented FMM, it is shown that adiabatic heterostructures designs are advantageous compared to abrupt heterostructures for minimizing the cavity scattering loss.
\end{abstract}

\begin{IEEEkeywords}
VCSEL, Vertical-cavity surface-emitting laser, HCG, High contrast grating, quality factor.
\end{IEEEkeywords}

\IEEEpeerreviewmaketitle

\section{Introduction}
\begin{figure}[t]
\centering
\includegraphics[width=0.45\textwidth]{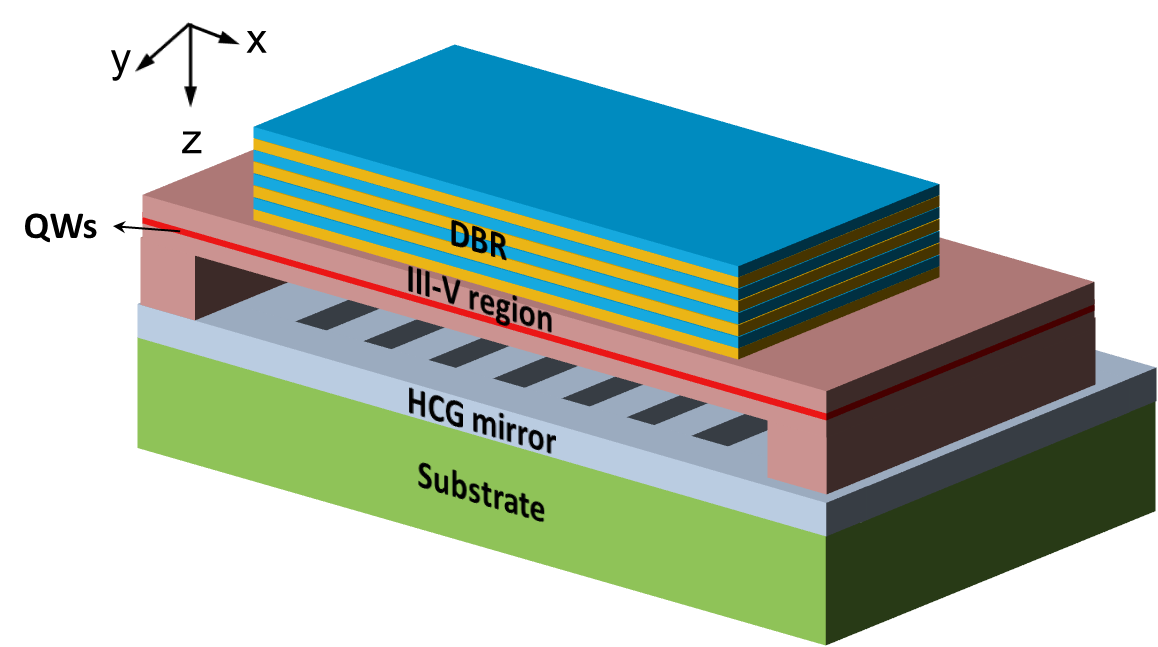}
\caption{Schematic view of a HCG-based vertical cavity structure with a top reflector realized by a distributed Bragg reflector (DBR) and a bottom reflector realized by a high-contrast grating (HCG).}
\label{fig:HCGCavitySch}
\end{figure}

\IEEEPARstart{V}{ertical}-cavity surface-emitting lasers (VCSELs) have been suggested as a candidate light source for near-future optical interconnects \cite{Miller2010, Dalir2013}. State-of-the-art VCSELs with distributed Bragg reflectors (DBRs) can be power-efficient \cite{Hofmann2012} and be integrated to silicon (Si) \cite{Haglund2015}. However, using the DBR-based VCSEL structures for chip-level optical interconnects is not feasible, since coupling output light from an vertical cavity into an in-plane Si waveguide is not straightforward, requiring extra efforts such as the preserve of a grating coupler. Recently, a hybrid III-V-on-Si vertical-cavity laser (hybrid VCL) structure has been proposed using a Si high-contrast grating (HCG) reflector, as shown in Fig. \ref{fig:HCGCavitySch} \cite{Chung2010}. Thanks to its capability of high-speed operation \cite{Park2015_APR} and in-plane light emission \cite{Viktorivitch2013, Park2015}, enabled by the HCG, the hybrid VCL appears promising for Si-integrated photonic circuit applications including chip-level optical interconnects. The HCG is a special class of subwavelength gratings, in which the grating period is close to the incident light wavelength and there is a large refractive-index contrast between the grating bars and surrounding materials\cite{Mateus2004}. Due to the large refractive-index contrast, HCGs can provide extraordinary properties that have not been reported for conventional guided-mode resonance (GMR) filters with similar periodicity, such as high reflectivity over a broad bandwidth \cite{Mateus2004, Magnusson2008}, and high quality-factor (Q-factor) resonances ($Q$$>$$10^7$) \cite{Zhou2008, Wu2013}. Employing the HCG in VCSELs as reflector maintains or even improves the desirable properties of DBR-based VCSELs such as a small modal volume, a low threshold current and a high power-efficiency \cite{Huang2007}. Moreover, it results in novel properties such as MEMS-based efficient wavelength tunability \cite{Huang2008, Chung2010a}, strong single-transverse-mode operation \cite{Huang2007, Chung2008}, and engineering of output beam profiles \cite{Fattal2010, Carletti2011}. Furthermore, it has been recently shown that in-plane heterostructure implemented in HCG-based vertical cavities \cite{Sciancalepore2011} enables exotic heterostructure configurations \cite{Taghizadeh2015a}, which are attractive for fundamental physics studies \cite{Wang2015}.

Even though there are several methods for simulating and modeling a HCG, the simulation of HCG-based vertical cavities, such as hybrid VCL structures, has been usually performed using the finite-difference time-domain (FDTD) method \cite{Chung2008, Chung2010, Chung2010a, Sciancalepore2011, Viktorivitch2013, Park2015}. FDTD simulations are powerful for analysis of many types of structures. However, long simulation times and a large amount of memory are required to simulate HCG-based cavities, since the entire simulation domain is large (tens of $\mu$m) while the smallest characteristic dimension is small (hundreds of nm). Recently, there have been several attempts to simulate HCG-based cavity structures using the other techniques \cite{Gebski2014, Tibaldi2015}. In this work, the Fourier modal method (FMM) is employed, since it is particularly efficient for solving Maxwell's equations in periodic structures. The FMM is a special class of mode expansion techniques (or modal methods). In modal methods, the structure is discretized into layers, the eigenmodes of each layer are determined, and the eigenmodes of adjacent layers are connected using mode matching at interfaces. Modal methods provide valuable insight about the physics of simulated structures by giving direct access to interesting physical parameters such as mode profiles, effective refractive-indices and scattering coefficients in the structure \cite{Lavrinenko2013}. HCG-based vertical cavities are layered structures and comprise a periodic grating layer, which makes them suitable for simulation using the FMM. Furthermore, in the FMM, several important parameters of interest such as reflectivity/transmissivity and cavity Q-factor are directly obtained without requiring any post-processing of simulation data, which is a strong advantage for numerical studies of HCG-based vertical cavities. Different numerical tweaks such as Li's factorization rule \cite{Li1996, Li1997}, and application of structural mirror-reflection symmetries \cite{Li2003} can be implemented to improve the performance of the FMM considerably. Furthermore, absorbing boundary layers or perfectly matched layers (PMLs), implemented as regions of non-linear complex stretched-coordinate, can efficiently reduce the effect of finite computational domain \cite{Hugonin2005}.  

In this paper, we report on our in-house simulator based on the FMM, developed for studies of HCG-based cavities, which introduces new features for achieving increased numerical efficiency. Using this simulator, hybrid VCL structures have been numerically studied with emphasis on the analysis of 3D structures. Firstly, the implemented FMM with PML method is introduced briefly. Several different numerical techniques for determining the resonance wavelength and Q-factor of a cavity mode are introduced and compared. Among them, the round-trip matrix approach appears very efficient for determining several transverse modes of the cavity at once. The numerical convergence of resonance wavelength and Q-factor is also investigated. The Q-factor is found to be much more difficult to accurately simulated than the resonance wavelength and this is explained. Furthermore, a method to simplify 3D simulations to several low-dimensional simulations is also suggested. The methods for computing the cavity dispersion, which is an important characteristic of the HCG-based cavities, is illustrated with a numerical example. Finally, the influence of adiabatic and abrupt in-plane heterostructures on the HCG-based cavity performance in transverse directions is investigated numerically. The reliability and numerical efficiency of the simulator presented in this work has been proven in our recent studies of HCG and related structures \cite{Taghizadeh2014b, Park2015, Taghizadeh2015a, Taghizadeh2014_HG, Taghizadeh2015_HG}.

\section{Fourier Modal Method}
\noindent The FMM, which is also referred to as the rigorous coupled wave analysis (RCWA), was introduced in early 80s to rigorously solve grating problems \cite{Moharam1981}. Due to its simplicity and robustness, it was used later for simulating other types of structures such as corrugated waveguides \cite{Lalanne2000, Silberstein2001}, optical couplers \cite{Silberstein2001}, plasmonics structures \cite{Weiss2011}, and photonic crystal (PhC) structures \cite{Botten2001, Lecamp2007, Lasson2014}. In this approach, by assuming a distinct propagation direction, the structure is discretized into layers. For each layer, the eigenmodes including both guided and radiated modes are calculated by solving an eigenvalue problem, which is obtained by expanding the eigenmodes on a basis set of exponential functions. Due to the linearity of Maxwell's equations, the total field in each layer is a summation over all of these eigenmodes. Continuity of transverse electric and magnetic field components at the layer interfaces results in a linear relationship between the mode coefficients in adjacent layers. Solving these equations for the mode coefficients, the field is determined for the entire structure.

\subsection{Conventions}
Here, the main physical and mathematical conventions employed in this work are reviewed. The main propagation direction is denoted by the $z$-direction and is referred to as the longitudinal direction, while the $x$-, $y$-directions are referred to as the transverse (or in-plane) directions (c.f. Fig. 1). For transverse electric (TE) or transverse magnetic (TM) polarization, the electric field is assumed to be parallel or perpendicular to the grating bars, respectively. All vectors and matrices are indicated by bold letters. Table \ref{tab:Conventions} shows a list of main conventions employed in this article.
\begin{table}
\caption{The list of main conventions employed in this work with their corresponding meaning.} 
\label{tab:Conventions}
\centering
\begin{tabular}{M{1.1cm} | M{4.8cm}} 
\hline
Symbol & Name \\  
\hline
$x,y,z$ & Cartesian coordinate \\
$\vec{\mat{r}}$ & Position vector \\ 
$\partial_\alpha$ & Derivative operator ($\partial/\partial \alpha$) \\ 
$\mat{I}$ & Identity matrix \\ 
$\mat{O}$ & Zero matrix \\
\hline
$\epsilon_0$ & Vacuum permittivity \\
$\mu_0$ & Vacuum permeability \\
$c$ & Speed of light in vacuum \\
\hline
$\omega$ & Angular frequency \\
$\lambda_0$ & Wavelength in vacuum \\
$k_0$ & Wavenumber in vacuum ($\omega/c=2\pi/\lambda_0$) \\
$\epsilon_r$ & Relative permittivity \\
$\mu_r$ & Relative permeability \\
$\vec{\mat{E}}$ & Electric field vector \\
$\vec{\mat{H}}$ & Magnetic field vector \\
\hline 
\end{tabular}
\end{table}

\subsection{Eigenmodes of a Layer}
After discretizing the structure into separate layers (the permittivity function of a layer has no spatial variation in the $z$-direction), the eigenmodes and their propagation constants for each layer are determined. Our starting point is Maxwell's equations for a source-free medium in the frequency domain, assuming $\exp(+j\omega t)$ as the time harmonic dependency. Only linear materials are considered here, i.e. $\vec{\mat{D}}=\epsilon_0\epsilon_r\vec{\mat{E}}$ and $\vec{\mat{B}}=\mu_0\mu_r\vec{\mat{H}}$. It is convenient to work with normalized magnetic field, i.e. $\vec{\tilde{\mat{H}}}=-j\sqrt{\mu_0/\epsilon_0}\vec{\mat{H}}$ and normalized coordinate system, i.e. $\vec{\tilde{\mat{r}}}=k_0 \vec{\mat{r}}$ (or equivalently $\tilde{\mat{\nabla}} \times=k_0^{-1} \mat{\nabla} \times$), since two similar equations are then obtained for the electric and magnetic fields as: 
\begin{subequations}
\label{eq:MaxwellCurls}
\begin{align}
\tilde{\mat{\nabla}} \times \vec{\mat{E}} &= \mu_r \vec{\tilde{\mat{H}}}, \\
\tilde{\mat{\nabla}} \times \vec{\tilde{\mat{H}}} &= \epsilon_r \vec{\mat{E}}.	
\end{align}
\end{subequations}

The PML is implemented as a complex coordinate stretching (which is a kind of coordinate transformation) similar to the implementation in the FDTD technique \cite{Chew1997}. Therefore, Maxwell's equations are rewritten with coordinate stretching functions $g_x$ and $g_y$ for the $x$- and $y$-directions \cite{Hugonin2005a}. By eliminating the $z$-component of the electric and magnetic fields, the following equations are obtained for the transverse field components \cite{Rumpf2011}:
\begin{subequations}
\label{eq:MaxwellOper}
\begin{align}
&\partial_z \matc{E_x}{E_y} = \pounds_{EH}\matc{\tilde{H}_x}{\tilde{H}_y} \\
&\partial_z \matc{\tilde{H}_x}{\tilde{H}_y} = \pounds_{HE}\matc{E_x}{E_y}
\end{align}
\end{subequations}
where $\pounds_{HE}$ and $\pounds_{EH}$ are the differential field operators \cite{Granet2002}:
\begin{subequations}
\label{eq:MaxwellOperPML}
\begin{align}
&\pounds_{EH} = \matd{-f_x\partial_x\epsilon_r^{-1}f_y\partial_y}{\mu_r+f_x\partial_x\epsilon_r^{-1}f_x\partial_x}{ -\mu_r-f_y\partial_y\epsilon_r^{-1}f_y\partial_y}{f_y\partial_y\epsilon_r^{-1}f_x\partial_x}, \\ 
&\pounds_{HE} = \matd{-f_x\partial_x\mu_r^{-1}f_y\partial_y}{\epsilon_r+f_x\partial_x\mu_r^{-1}f_x\partial_x}{-\epsilon_r-f_y\partial_y\mu_r^{-1}f_y\partial_y}{f_y\partial_y\mu_r^{-1}f_x\partial_x}, 
\end{align}
\end{subequations}
where $f_x=(\partial g_x/\partial{{x}})^{-1}$ and $f_y=(\partial g_y/\partial{{y}})^{-1}$. The coordinate stretching functions are unity transformations outside the PMLs, i.e. $g_x=x$ and $g_y=y$, respectively and they are complex-value functions inside the PMLs, with the imaginary part of the transformation damping the outgoing waves \cite{Pisarenco2010}. Here, the coordinate stretching function suggested in Ref. \cite{Hugonin2005} is employed. 

In the optical frequency range and for most practical materials, $\mu_r$ is unity \cite{Joannopoulos2008}. For each layer, the permittivity function is invariant in the $z$-direction, i.e. $\epsilon_r(x,y,z)=\epsilon_r(x,y)$. In the FMM, it is assumed that the structure is periodic in both $x$- and $y$-directions with periodicities of $\Lambda_x$ and $\Lambda_y$, respectively. Therefore, the permittivity $\epsilon_r$ and coordinate stretching factors $f_x$ and $f_y$ can be expanded as a 2D Fourier series where each term corresponds to a spatial harmonic. Furthermore, due to Bloch's theorem, a pseudo-periodic Fourier series can be used for expanding the $\vec{\mat{E}}$ and $\vec{\tilde{\mat{H}}}$ fields. These Fourier series expansions transform Maxwell's differential equations to their matrix representations as \cite{Rumpf2011}: 
\begin{subequations}
\label{eq:MaxwellMatrix}
\begin{align}
&\partial_z \matc{\mat{S}_x}{\mat{S}_y} = \mat{P}\matc{\mat{U}_x}{\mat{U}_y}, \\ 
&\partial_z \matc{\mat{U}_x}{\mat{U}_y} = \mat{Q}\matc{\mat{S}_x}{\mat{S}_y}, 
\end{align}
\end{subequations}
where the vectors $\mat{S}_x$, $\mat{S}_y$, $\mat{U}_x$ and $\mat{U}_y$ contain the Fourier series coefficients of the transverse field components $E_x$, $E_y$, $\tilde{H}_x$ and $\tilde{H}_y$, respectively. Matrices $\mat{P}$ and $\mat{Q}$ are:
\begin{subequations}
\label{eq:MaxwellMatrixPQ}
\begin{align}
&\mat{P} = \matd{\mat{F}_x\mat{K}_x\mat{\mathcal{E}}_z^{-1}\mat{F}_y\mat{K}_y}{\mat{I}-\mat{F}_x\mat{K}_x\mat{\mathcal{E}}_z^{-1}\mat{F}_x\mat{K}_x}{\mat{F}_y\mat{K}_y\mat{\mathcal{E}}_z^{-1}\mat{F}_y\mat{K}_y-\mat{I}}{-\mat{F}_y\mat{K}_y\mat{\mathcal{E}}_z^{-1}\mat{F}_x\mat{K}_x}, \\
&\mat{Q} = \matd{\mat{K}_x\mat{K}_y}{\mat{\mathcal{E}}_y-\mat{K}_x^2}{\mat{K}_y^2-\mat{\mathcal{E}}_x} {-\mat{K}_y\mat{K}_x}, 
\end{align}
\end{subequations}	
where $\mat{F}_x$ and $\mat{F}_y$ are the Toeplitz matrices defined from the Fourier series coefficients for the functions $f_x$ and $f_y$, respectively \cite{Hugonin2005a}, and matrices $\mat{\mathcal{E}}_x$, $\mat{\mathcal{E}}_y$ and $\mat{\mathcal{E}}_z$ are obtained from $\epsilon_r$ as discussed in Ref. \cite{Li1997}. $\mat{K}_x$ and $\mat{K}_y$ are diagonal matrices containing the normalized spatial wavevector components ${k}_{xm}=m\lambda/\Lambda_x$ and ${k}_{yn}=n\lambda/\Lambda_y$, respectively. $m$ and $n$ are the Fourier series indices, $m=-m_x,\cdots,-1,0,1,\cdots m_x$ and $n=-n_y,\cdots,-1,0,1,\cdots n_y$, and the total number of Fourier terms are $M_x=2m_x+1$, $N_y=2n_y+1$ and $N_t=M_xN_y$ in the $x$-direction, $y$-direction and both directions, respectively. All the matrices in Eqs. (\ref{eq:MaxwellMatrixPQ}a) and (\ref{eq:MaxwellMatrixPQ}b) are squared matrices with the size of $N_t$.

By combining these two equations, an expression for $\{\mat{S}_x,\mat{S}_y\}$ is obtained, and matrix $\mat{\Omega}$ is defined as $\mat{\Omega}=\mat{P}\mat{Q}$. Assuming a $z$-dependency in the form of $\exp{({\gamma} {z})}$ for the field components, the differential equations are transformed to an eigenvalue problem for the matrix $\mat{\Omega}$, which is solved easily by a standard eigenvalue software package (MATLAB in this work): 
\begin{equation}
\partial_z^2 \matc{\mat{S}_x}{\mat{S}_y} = \mat{\Omega} \matc{\mat{S}_x}{\mat{S}_y} \quad \Rightarrow \quad \mat{{\gamma}}^2 \matc{\mat{S}_x}{\mat{S}_y} = \mat{\Omega} \matc{\mat{S}_x}{\mat{S}_y}.
\end{equation}
After solving the eigenvalue problem, eigenvalues $b_i$ ($i$=1,2,$\cdots$,$2N_t$) are found, where $b_i={\gamma}_i^2$. Since, $\gamma_i=\pm \sqrt{b_i}$, there are two eigenvalues which share a single eigenvector, and the appropriate sign is chosen as it is discussed in Refs. \cite{Granet1996, Hench2008}. Each ${\gamma}_i$ is referred to as an eigenmode propagation constant and its corresponding eigenvector $(\mat{w}_{x,i},\mat{w}_{y,i})$ as an electric field eigenmode. All the eigenvectors  form a matrix $\mat{W}$, where eigenvector $(\mat{w}_{x,i},\mat{w}_{y,i})$ is the $i$th column. A diagonal matrix $\mat{\Gamma}$ is also defined, where $\gamma_i$ are the diagonal values, and consequently $\mat{\Omega}\mat{W}=\mat{W}\mat{\Gamma}^2$. Furthermore, using Eqs. (\ref{eq:MaxwellMatrix}a) and (\ref{eq:MaxwellMatrix}b), the magnetic field eigenmode $(\mat{v}_{x,i},\mat{v}_{y,i})$ is the $i$th column of the matrix $\mat{V}=\mat{Q}\mat{W}\mat{\Gamma}^{-1}$.

The total field in a layer is a weighted summation of all the eigenmodes with their corresponding propagation constants \cite{Rumpf2011}:
\begin{subequations}
\label{eq:ch2_TotalFieldLayer}
\begin{align}
&\matc{\mat{S}_x}{\mat{S}_y} = \mat{W}\exp{[-\mat{\Gamma}({z}-{z}_1)]}\mat{c}_p + \mat{W}\exp{[\mat{\Gamma}({z}-{z}_2)]}\mat{c}_n, \\
&\matc{\mat{U}_x}{\mat{U}_y} = -\mat{V}\exp{[-\mat{\Gamma}({z}-{z}_1)]}\mat{c}_p + \mat{V}\exp{[\mat{\Gamma}({z}-{z}_2)]}\mat{c}_n. 
\end{align}
\end{subequations}
where $\mat{c}_p$ and $\mat{c}_n$ are the expansion coefficient vectors of the propagating waves in positive and negative directions, respectively, and the layer starts at ${z}={z}_1$ and ends at ${z}={z}_2$. The phase factors, $\exp{[-\mat{\Gamma}z_1]}$ and $\exp{[-\mat{\Gamma}z_2]}$, are introduced to prevent numerical overflow as explained in Ref. \cite{Granet1996}. For a given excitation field, $\mat{c}_p$ and $\mat{c}_n$ can be found by employing appropriate boundary conditions as discussed in section \ref{sec:SMatrices}.

\subsection{Structural Mirror Symmetries}
If the permittivity function is an even function of one of the coordinates, the solution to Maxwell's equations will be either even or odd in that coordinate \cite{Joannopoulos2008}. Considering only the mirror symmetries of the permittivity function in the transverse directions, four cases are possible for a structure; no symmetry, $x$-axis mirror symmetry, $y$-axis mirror symmetry, and simultaneous $x$- and $y$-axis mirror symmetries. In the FMM, the mirror symmetries in real space are transferred to the Fourier space \cite{Li2003}, and they can be employed to considerably reduce the simulation effort, i.e. time and memory. With no use of mirror symmetries, the size of the eigenvalue problem of $\mat{\Omega}$ is $2N_t$ and the total number of matrix elements is $4N_t^2$. If the structure has mirror symmetry in both transverse directions, the matrix size can be reduced to one fourth. Since the eigenvalue computation time is roughly proportional to the third power of the matrix size, employing the two mirror symmetries results in a factor of $4^3$ ($=64$) reduction of the computation time. Moreover, the number of matrix elements is reduced by a factor of $4^2$ ($=16$). The method introduced in Refs. \cite{Li2003, Zhou2004} is implemented in this work to take advantage of the mirror symmetries.

\subsection{Homogeneous Layer Eigenmodes}
For a homogeneous layer, i.e. a layer with no variation of the relative permittivity function in transverse directions, the eigenmodes of the layer can be obtained analytically, provided that no PML is included. These eigenmodes correspond to the Rayleigh expansions for diffraction orders \cite{Li1997}. If the PML is implemented, the eigenmodes can not be found analytically anymore. However, if there are several homogeneous layers, a simple relationship can be found between their eigenmodes, as discussed in Ref. \cite{Yala2009} for 2D problems. Here, we derive a similar relation for 3D cases. Assume that the eigenvalue problem is solved for free-space and its matrices for the eigenmode electric field and propagation constant are $\mat{W}_0$ and $\mat{\Gamma}_0$, respectively. The eigenvalue problem for determining the eigenmodes of layer with permittivity of $\epsilon_r$ has a solution of ($\mat{W}$, $\mat{\Gamma}$), where
\begin{subequations}
\begin{align}
&\mat{W} = \mat{W_0}, \\
&\mat{\Gamma}^2 = \mat{\Gamma}_0^2+(1-\epsilon_r)\mat{I}. 
\end{align}
\end{subequations}

\subsection{Two-Dimensional Limit}
If the structure is invariant in one direction, for instance in the $y$-direction, the 3D problem can be simplified to two individual 2D problems for the TE and TM polarizations. For the TE polarization, the non-zero field components are $(E_y,\tilde{H}_x,\tilde{H}_z)$, while for the TM polarization only $(E_x,E_z,\tilde{H}_y)$ are non-zero. The matrix size for a 2D problem is reduced to half, which results in an eight-fold and four-fold reduction of computational time and required memory, respectively. The expressions for the $\mat{P}$ and $\mat{Q}$ matrices for the two polarizations can be obtained simply by letting $\mat{K}_{y}=\mat{O}$:
\begin{subequations}
\begin{align}
&\mat{\Omega} = \mat{P}\mat{Q} = \matd{\mat{\Omega_{11}}}{\mat{O}}{\mat{O}}{\mat{\Omega_{22}}}, \\
&TM: \mat{\Omega_{11}} = ( \mat{F}_{x}\mat{K}_{x}\mat{\mathcal{E}}_{z}^{-1}\mat{F}_{x}\mat{K}_{x}-\mat{I}) \mat{\mathcal{E}}_{x}, \\
&TE: \mat{\Omega_{22}} = ( \mat{F}_{x}\mat{K}_{x}\mat{F}_{x}\mat{K}_{x}-\mat{\mathcal{E}}_{y}).
\end{align}
\end{subequations}
Since $\mat{\Omega}$ is a block diagonal matrix, its eigenvalues are a combination of the eigenvalues of its block matrices $\mat{\Omega}_{11}$ and $\mat{\Omega}_{22}$, which correspond to the TM and TE polarization problems, respectively. Without the PML implementation ($\mat{F}_{x}=\mat{I}$), these expressions will be similar to the expressions derived in Refs. \cite{Moharam1995, Lalanne1996}. However, the eigenvalue problem for the TM polarization is found based on the electric field $E_x$ (instead of the magnetic field $\tilde{H}_y$).

\subsection{Scattering Matrices \label{sec:SMatrices}}
In order to connect the fields in neighboring layers, the continuity of transverse field components, i.e. $E_x$, $E_y$, $\tilde{H}_x$ and $\tilde{H}_y$, at the layer interfaces should be applied. It results in a linear relationship between the field coefficients, $\mat{c}_p$ and $\mat{c}_n$ in two neighboring layers. In a multilayer structure, various algorithms can be used to connect the coefficients, such as transfer matrices \cite{Moharam1995a}, hybrid matrices \cite{Tan2006}, admittance matrices \cite{Dems2007a}, and scattering matrices (S-matrices) \cite{Li1996, Moharam2004, Rumpf2011}. Among them, the S-matrices are used widely in the literature, due to various advantages such as elegant physical interpretations, unconditional stability and memory efficiency \cite{Rumpf2011}. There are several variants of S-matrix implementations for the FMM \cite{Li1996, Rumpf2011}. In the variant implemented in this work, all the layers are separated by a gap composed of free-space \cite{Moharam2004, Rumpf2011}. As long as the free-space gap thickness is zero, it has no influence on the performance of the structure. The advantage of using the artificial free-space regions, is that the S-matrices of each layer only depends on the layer itself, not on its adjacent layers \cite{Moharam2004, Rumpf2011}. It reduces the numerical effort for the cases where only one layer is changing, since only the S-matrices of that layer should be updated instead of all adjacent layers.

\section{Resonance Wavelength and Q-Factor \label{sec:QFactorMethods}}
\noindent The important characteristics of a resonant optical cavity, are the resonance wavelengths $\lambda_r$ and Q-factors $Q$ of the cavity modes. In this section, several methods for calculating the resonance wavelength and Q-factor, using the FMM, are compared. The advantages and disadvantages of these methods are discussed in section \ref{sec:QFactorExample}.

\subsection{Quasi-Normal Mode Approach}
Since an optical cavity is leaky, it forms an open system, and the eigenmodes of Maxwell's equations have complex frequencies \cite{Settimi2010, Kristensen2013}. These eigenmodes are referred to as quasi-normal modes (QNMs). Using the concept of QNMs provides a convenient framework for working with optical cavities, since the real parts of QNM eigenvalues correspond to the position and the imaginary parts of QNM eigenvalues correspond to Q-factor of the resonances in the spectral transmissivity \cite{Severini2005}. In the QNM picture, the optical cavity is viewed as a passive open system with only emission out of the cavity and no waves incident to the cavity. Therefore, outgoing boundary conditions should be satisfied for the QNM. In the FMM, the outgoing boundary condition is satisfied in the $z$-direction \cite{Lasson2014}. Using the concept of S-matrices, QNMs can be found as the non-trivial solution of the following equation for $\mat{c}_{in}$ approaching zero \cite{Weiss2011}:
\begin{equation}
\mat{c}_{out} = \mat{S}\mat{c}_{in} \quad \rightarrow \quad \mat{S}^{-1}\mat{c}_{out} = 0,
\end{equation} 
which corresponds to the poles of the $\mat{S}^{-1}$ matrix. In other words, the QNM complex frequencies are those that make the determinant of the $\mat{S}^{-1}$ matrix zero. There is an equivalent method, referred to as the round-trip matrix method, which is easier and more efficient to implement in the FMM compared to the first method \cite{Lasson2014}. In this approach, the QNMs are found as the eigenmodes of a round-trip matrix $\mat{U}$ inside the cavity with an eigenvalue of 1. They are found for a complex frequency $\tilde{\omega}$, and the QNMs frequency will be the real part of $\tilde{\omega}$, and their Q-factor are obtained by $Q=\Re(\tilde{\omega})/[2 \Im(\tilde{\omega})]$ \cite{Lecamp2005, Lasson2014}. The round-trip matrix $\mat{U}$ is obtained at an arbitrary plane inside the cavity by determining the S-matrices seen to the left $\mat{S}_{11}^{(left)}$ and right $\mat{S}_{11}^{(right)}$: 
\begin{equation}
\mat{U} = \mat{S}_{11}^{(right)}\mat{S}_{11}^{(left)}.
\end{equation}

\subsection{Fabry-Perot Approach}
In this approach, the round-trip matrix method is employed in real frequency domain instead of complex frequency plane of the QNM approach, but the phase of eigenvalue of matrix $\mat{U}$ is considered  \cite{Gregersen2010}. If $R_r$ is the eigenvalue of matrix $\mat{U}$, the resonance wavelength $\lambda_r$ is found as the wavelength, which makes the phase of $R_r$ zero, i.e. $arg(R_r)$=0. This is equivalent to a constructive interference condition for a complete round-trip inside the cavity. The Q-factor is then obtained as \cite{Gregersen2010}:
\begin{equation}
\label{eq:QFactorII}
Q = \frac{\lambda_r}{2(1-R_r)} \frac{\partial}{\partial \lambda} arg(R_r)
\end{equation} 
where the derivative is evaluated at $\lambda_r$ where $arg(R_r)$=0.

\subsection{Spectral Approach}
The most straightforward method for determining the cavity resonances is to consider the reflectivity (or transmissivity) spectrum of an optical cavity, i.e. consider the response to an incident wave. In this method, a notch in the reflectivity spectrum (or a peak in transmissivity spectrum) corresponds to a resonance wavelength. The Q-factor is estimated as the ratio of the resonance wavelength $\lambda_r$ and the full-width half-maximum (FWHM) bandwidth of the notch (or peak), $Q=\lambda_r/\Delta \lambda$. Regarding the incident wave, one need to ensure that the incident wave can excite the cavity mode, e.g. the incident wave is required to have the same parity as the cavity mode profile.

\section{Low-Dimensional Simulations}

\begin{figure}
\centering
\includegraphics[width=0.35\textwidth]{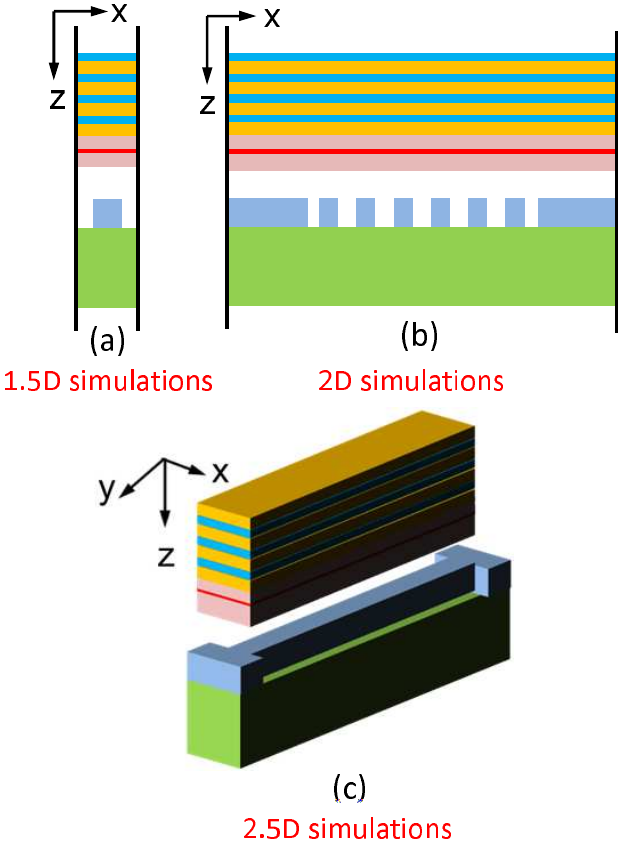}
\caption{Schematic view of a HCG-based vertical cavity structure used for low-dimensional simulations. (a) 1.5D simulations, (b) 2D simulations, and (c) 2.5D simulations.}
\label{fig:LowDimensionSch}
\end{figure}

\noindent Similar to other numerical techniques, it is highly demanding in terms of time and memory for the FMM to solve Maxwell's equations for a 3D structure. Thus, it remains an important challenge to develop techniques for reducing the computational effort. For a 3D HCG-based vertical cavity, we will show that it is possible to analyze the structure by considering lower-dimensional structures, as illustrated in Fig. \ref{fig:LowDimensionSch}. Thus, by performing low-dimensional simulations, one can estimate the cavity loss in each transverse direction separately. This approach is particularly useful for optimizing the in-plane heterostructure design, leading to a huge speed improvement, compared to full 3D simulations. Firstly, a so-called 1.5-dimensional (1.5D) simulation is carried out, where  only one period of the HCG is used, and the structure is assumed to be uniform in the $y$-direction. Indeed, it is a two-dimensional (2D) simulation for a structure with an infinite number of grating periods due to the periodic boundary condition in the $x$-direction. This 1.5D simulation can be employed to compute the reflectivity, transmissivity and cavity dispersion. For a 2D simulation, a finite number of grating periods is considered as illustrated in Fig. \ref{fig:LowDimensionSch}(b), and the structure is assumed to be uniform in the $y$-direction, similar to the 1.5D case. Absorbing boundary conditions are implemented in the $x$-direction. A 2D simulation can be employed to estimate the loss in the $x$-direction, and also to design a heterostructure in this direction. Figure \ref{fig:LowDimensionSch}(c) shows a structure used for 2.5-dimensional (2.5D) simulations, in which only a single grating period is considered, similar to the 1.5D structure. However, the structure is not uniform in the $y$-direction any more, and absorbing boundary conditions are employed in this direction. A 2.5D simulation can be utilized to design an in-plane heterostructure in the $y$-direction.

The photon life-time of an optical cavity quantifies the decay rate of the energy from the cavity, and is related to the Q-factor of the cavity as $\tau=Q/\omega$. Photons can escape from the cavity in any direction. Assuming a box surrounding the cavity structure, three photon life-times $\tau_x$, $\tau_y$ and $\tau_z$ are defined, which quantifies the cavity decay rates corresponding to different propagation directions, in the $x$-, $y$- or $z$-directions, respectively. If we assume the cavity losses in different directions are not correlated, a simple relationship is obtained for $\tau$ or equivalently the Q-factor as: 
\begin{subequations}
\label{eq:DirectionQFactor}
\begin{align}
&\frac{1}{\tau}=\frac{1}{\tau_x}+\frac{1}{\tau_y}+\frac{1}{\tau_z}, \\
&\frac{1}{Q}=\frac{1}{Q_x}+\frac{1}{Q_y}+\frac{1}{Q_z}, 
\end{align}
\end{subequations}
where $Q_i=\omega\tau_i$ with $i$=$x,y,z$ are the Q-factors corresponding to the each direction. For a HCG-based cavity, $Q_z$, $Q_x$ and $Q_y$ can be estimated by performing 1.5D, 2D and 2.5D simulations of the structure, respectively. Furthermore, the resonance wavelength of the cavity can be estimated using the cavity dispersion as discussed in Ref. \cite{Taghizadeh2015a}. Therefore, it is possible to estimate the resonance wavelength and Q-factor of a cavity mode without performing a full 3D simulation.

\section{Numerical Examples}
\noindent In this section, we provide several numerical simulation examples of HCG-based vertical cavities. The considered structure is the hybrid VCL structure for optical interconnect applications at 1550-nm wavelength reported in Ref. \cite{Park2015}. It consists of a dielectric DBR, an InP active region, and a grating layer formed in the Si layer of an Si-on-insulator (SOI) wafer. The parameters for the structure are summarized in Table \ref{tab:LayerThickness}. The HCG is designed to be highly reflective (above 99.9\%) for TM-polarized light at 1550-nm wavelength, and having a bandwidth larger than 50 nm. The thickness of the air gap layer between the III-V and low-index gap layer is designed to result in constructive interference, considering the reflection phase of the HCG. If PMLs are required for simulations, the thicknesses are chosen as 0.5 and 2 $\mu$m in the $x$- and $y$-directions, respectively. If not stated otherwise, the grating layer comprises 14 grating bars with a period of 640 nm and a bar width of 371.2 nm, and is terminated to Si in the $x$-direction. The grating bars are 6 $\mu$m long in the $y$-direction, where the last 3 $\mu$m section has a smaller width of 315.5 nm to form a heterostructure in the $y$-direction (c.f. Fig. 6). Firstly, the convergence behavior of the resonance wavelength and Q-factor of a cavity mode is investigated in both transverse directions, and the difference in the convergence rate of these two quantities is discussed. Secondly, three different numerical techniques for calculating the Q-factor are compared. Next, the validity of low-dimensional simulations is tested for the example HCG-based cavity. Finally, a vertical cavity with an in-plane heterostructure is designed and investigated by using the low-dimensional simulation technique. 

\begin{table}
\renewcommand{\arraystretch}{1.2} 
\caption{Layer refractive-indices and thicknesses used in simulations. }
\label{tab:LayerThickness}
\centering
\begin{tabular}{M{1.4cm} | M{1.5cm} | M{1.9cm} | M{2.1cm}} 
\hline
Layer name & Refractive index & Thickness & Comment \\  \hline
Superstrate & $n_{sup}$=1 & $t_{sup}=\infty$ & Air infinite half space \\ \hline
DBR-h & $n_{h}$=3.48 & $t_h$=111.4 nm & 4-pairs \\ 
DBR-l & $n_{l}$=1.48 & $t_l$=261.8 nm & Si/SiO$_2$ DBR \\ \hline
III-V & $n_{a}$=3.166 & $t_a$=704.4 nm & InP active region \\ \hline
Low-index gap & $n_{c}$=1.0 & $t_c$=704.4 nm & Air \\ \hline
Grating & $n_h$=3.48, $n_l$=1.0 & $t_g$=430 nm & Si grating \\ \hline
Substrate & $n_{sub}$=1.48 & $t_{sub}=\infty$ & SiO$_2$ infinite half space \\ \hline
\end{tabular}
\end{table}

\subsection{Convergence Rates}
To confirm the validity of our simulator based on the FMM, the convergence behavior has been investigated by evaluating the resonance wavelength and Q-factor parameters as a function of the total number of Fourier terms $N_t$ (=$M_xN_y$). Numerically, if the relative error of the desired parameter is within the acceptable error, we say that the simulation has converged. In this work, the relative error of a parameter is defined with respect to its previous value, which is obtained for the nearest and smaller $N_t$. By performing complete 3D simulations of a HCG-based cavity, it is practically impossible to determine the $M_x$ and $M_y$ values, which guarantee convergence, due to the very long simulation time required. However, by employing the low-dimensional simulations, one can estimate the required number of Fourier terms in the $x$- and $y$-directions separately, and use those numbers for a subsequent 3D simulation. Therefore, we investigate the convergence of resonance wavelength and Q-factor of the fundamental mode for the example HCG-based cavity in 2D and 2.5D simulations as a function of number of Fourier terms in the $x$-direction, $M_x$, and $y$-direction, $N_y$, respectively. 

\begin{figure}
\centering
\includegraphics[width=0.48\textwidth]{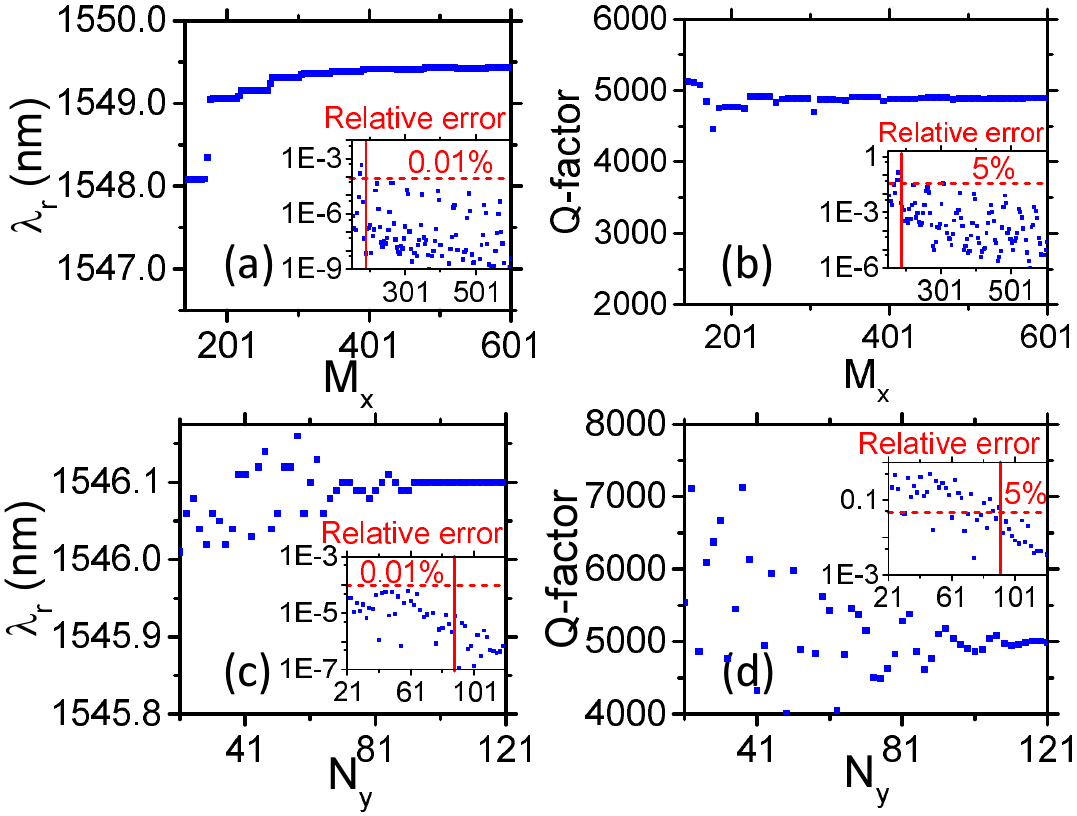}
\caption{(a),(c) Resonance wavelength, and (b),(d) Q-factor of the fundamental mode for the example HCG-based vertical cavity in (a),(b) 2D simulations versus $M_x$, and (c),(d) 2.5D  versus $N_y$. The insets show the relative error. The uncertainty of the resonance wavelength is approximately three orders of magnitude smaller than for the Q-factor. If the number of Fourier terms is larger than $M_x$=185 and $N_y$=87 in the $x$- and $y$-directions respectively, the uncertainty in resonance wavelength and Q-factor value will be less than 0.01 and 5 \%, respectively.}
\label{fig:ConvergenceTest}
\end{figure}

Figures \ref{fig:ConvergenceTest}(a) and (b) illustrate the convergence test of the resonance wavelength and Q-factor of the fundamental cavity mode for 2D simulations. Similarly, Figs. \ref{fig:ConvergenceTest}(c) and (d) show similar graphs in 2.5D simulations. In both 2D and 2.5D simulations, the convergence rate for Q-factor is lower than that of the resonance wavelength by approximately three orders of magnitude, which has been also reported for other numerical techniques, e.g., FDFD \cite{Ivinskaya2011}. This can be explained by using the analytic expression for the Q-factor in Fabry-Perot cavities \cite{Coldren2012}:
\begin{equation}
\label{eq:FPQFactor}
Q=-\frac{2\pi n_{c,g} t_\text{eff}}{\lambda_r}\frac{1}{\ln(r_1 r_2)},
\end{equation}
where $t_\text{eff}$ is the effective cavity length considering the mirror penetration depths, $n_{c,g}$ is the group refractive index of the cavity, and $r_1$ and $r_2$ are the reflectivity amplitudes for the two cavity mirrors. If the only uncertainties are in the value of $r_1$ and $r_2$, the uncertainty in the mode Q-factor will be:
\begin{equation}
\frac{\Delta Q}{Q}=-Q\Big(\frac{\Delta r_1}{r_1}+\frac{\Delta r_2}{r_2}\Big).
\end{equation}
Therefore, the relative error of the Q-factor is approximately Q-times higher than that of the reflectivity amplitude. On the other hand, the resonance wavelength is determined by the constructive interference condition as $\lambda_r=4\pi n_c t_c/(\phi_1+\phi_2-2m\pi)$, where $m$ is the longitudinal mode order, and $\phi_1$ and $\phi_2$ are the reflectivity phases of mirrors. Using this expression, the uncertainty of the resonance wavelength is on the same order as that of the mirror reflectivity phases. Since the reflectivity phase and amplitude have a similar convergence rate, the Q-factor shows a convergence rate which is lower by its value, compared to the convergence rate of resonance wavelength. Practically, if an uncertainty of 5\% for the Q-factor is acceptable, $M_x$=185 and $N_y$=87 are the minimum required number of Fourier terms in the $x$- and $y$-directions, respectively. It should be noted that the number of Fourier terms should be increased for an increasing number of grating bars or lengths of the grating bars, in order to obtain a specific relative error in the calculations. For a grating comprising $N_g$ grating bars, if $M_x$ is chosen to be $14N_g+$1 the calculation errors empirically does not exceed 5\% and 0.01\% for the Q-factor and resonance wavelength of cavity mode, respectively. 

\subsection{Q-factor Comparison \label{sec:QFactorExample}}
\begin{figure}[t]
\centering
\includegraphics[width=0.35\textwidth]{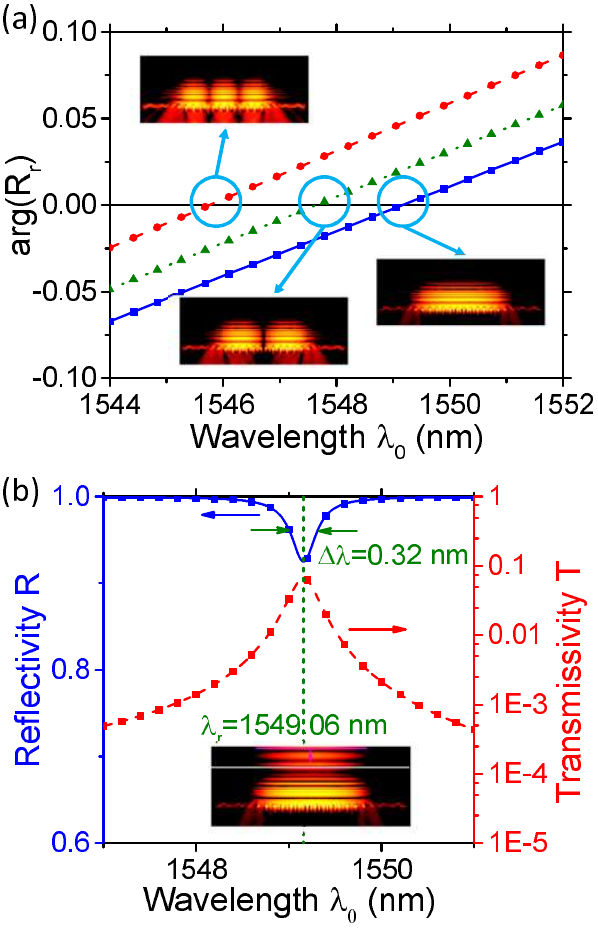}
\caption{(a) Evolution of eigenvalue phase of the round-trip matrix $\mat{U}$ as a function of wavelength for the example HCG-based cavity (2D simulation). Each line corresponds to a transverse mode which is also shown as an inset, and the resonance wavelength is at its intersection with zero phase line. (b) Reflectivity spectrum in linear scale (blue solid-line) and transmissivity spectrum in log scale (red dashed-line) of the example HCG-based cavity to an incident Gaussian beam with 6 $\mu$m beam waist width. The resonance wavelength is at $\lambda_r$=1549.06. All the insets field profiles are normalized $|H_y|$ of the structure (in dB scale).}
\label{fig:QFactor}
\end{figure}

In this section, the Q-factor and resonance wavelength of the fundamental mode for the example HCG-based cavity are computed with the three methods explained in section \ref{sec:QFactorMethods}. Although, for this comparison 2D simulations are performed here due the much less required simulation effort, 3D simulations will lead to a similar conclusion. For the QNM approach, the modes are determined by iterating the frequency in the complex plane until the eigenvalue of the matrix $\mat{U}$ become unity. The resonance wavelength is usually found in less than 10 iterations. But depending on the initial guess for resonance wavelength and the cavity Q-factor, more iterations may be required. For the Fapry-Perot approach, the eigenvalues of the round-trip matrix $\mat{U}$ are calculated as a function of the real frequency. For the implementation, the eigenvalues $R_r$ are computed for several wavelengths close to the estimated resonance wavelength. The resonance wavelength can be then found by interpolation or extrapolation of the eigenvalue phase, to the wavelength where $arg(R_r)$ becomes zero. Figure \ref{fig:QFactor} shows the phase of several eigenvalues of matrix $\mat{U}$ as a function of wavelength. Each line corresponds to a transverse mode, and its intersection with zero gives the resonance wavelength of that mode. The corresponding mode profile is shown in the inset. Finally, for the spectral approach, the fundamental mode of the HCG-based cavity is excited by a Gaussian wave from the superstrate. The reflectivity and transmissivity spectra are shown in Fig. \ref{fig:QFactor}(b) and the normalized field profile at the resonance wavelength is also shown in the inset.

Table \ref{tab:QFactorCompare} compares the simulation results from the three methods. The two methods employing the round-trip matrix $\mat{U}$, result in approximately similar values for both the resonance wavelength and Q-factor (their relative difference is less than 10$^{-7}$ for the resonance wavelength and 10$^{-3}$ for the Q-factor). These negligible differences can be attributed to the round-off errors in the calculations and is less than the uncertainty of the computations. The spectral approach provides a slightly different values compared to other methods, especially for the Q-factor, since the exact value of the FWHM bandwidth is difficult to evaluate for a large Q-factor mode. 

Although these three methods lead to approximately the same value for the cavity resonance wavelength and Q-factor, they are different from the numerical point of view. The main advantage of the QNM approach is that it can determine the resonance with an inaccurate initial guess for the resonance wavelength. However, this method can only find a single cavity mode for each function call. If there are several closely-spaced resonances, Fabry-Perot approach is much more efficient, since it can find all the resonance wavelengths and their Q-factor in one function call. Therefore, the Fabry-Perot method is suitable for determining several transverse modes of a HCG-based cavity. The spectral method takes the longest time, since it requires a wavelength scan with a fine resolution, particularly for a high Q-factor mode. Furthermore, the incident wave should be able to excite the mode efficiently, which could be difficult for modes with unknown or unusual spatial profiles. 

\begin{table}
\renewcommand{\arraystretch}{1.3}
\caption{Resonance wavelength $\lambda_r$ and Q-factor $Q$ of the fundamental mode for the example HCG-based cavity obtained from three different methods.}
\label{tab:QFactorCompare}
\centering
\begin{tabular}{c | c | c | c} 
\hline
Method & QNM & Fabry-Perot & Spectral \\ 
\hline 
$\lambda_r$ (nm) & 1549.0622 & 1549.0621 & 1549.06 \\ 
$Q$ & 4775.96 & 4781.52 & 4840 \\ \hline
\end{tabular}
\end{table}

\subsection{Cavity Dispersion}
Recently, it has been shown that the dispersion of a vertical cavity, i.e., the relation between the frequency $\omega$ and in-plane wavevector components $k_x$ or $k_y$ of a cavity mode, can be engineered by using the HCG as reflector and designing the dispersion of the HCG \cite{Wang2015, Taghizadeh2015a, Taghizadeh2015b}. The dispersion curvature is the second-order derivative of the frequency of a propagating mode with respect to the in-plane wavevector, and its inverse can be interpreted as an effective photon mass along the wavevector direction. As discussed in Ref. \cite{Taghizadeh2015a}, the dispersion curvatures along transverse directions can be engineered in HCG-based vertical cavities to have a specific positive, zero, or negative value. The control of dispersion characteristics opens a way to engineer the enhancement of the spontaneous emission through the Purcell factor \cite{Purcell1946} or control the properties of polariton lasers \cite{Zhang2015}. Furthermore, the freedom of engineering the cavity dispersion or equivalently the photon effective mass enables us to realize interesting in-plane heterostructures 
\cite{Taghizadeh2015a}, and plays an important role in controlling the importance of disorder effects as shown for photonic crystal lasers \cite{Weiqi2016}. Here, the method for rigorous calculation of the cavity dispersion is shown for the example HCG-based vertical cavity. For dispersion calculation 1.5D simulation is performed, since a unit cell of the structure is required for computing the dispersion.
 
\begin{figure}[t]
\centering
\includegraphics[width=0.48\textwidth]{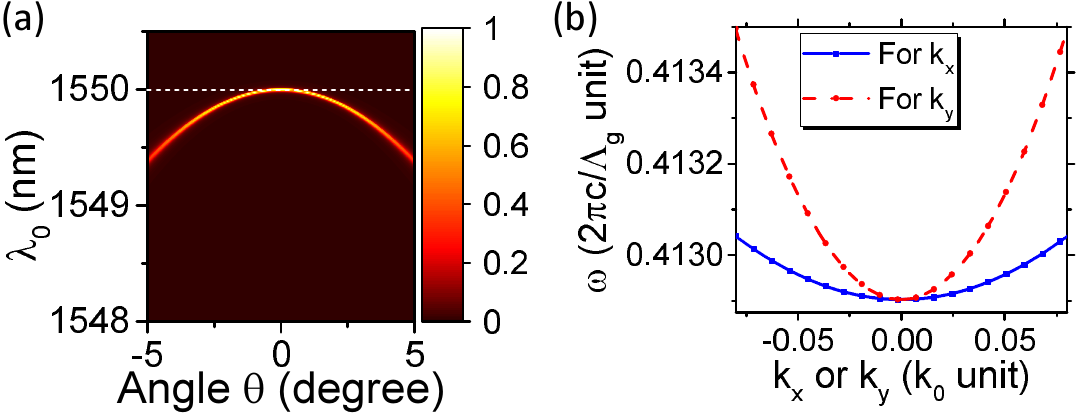}
\caption{(a) Contour maps of the transmissivity (color scale) versus incident angle $\theta$ and wavelength $\lambda_0$ for the example HCG-based vertical cavity to a TM-polarized plane-wave. The incident angle is defined with respect to the normal direction to the grating surface in the $x$-direction. The white dashed-line shows designed resonance wavelength of 1550 nm wavelength. The bright line in the contour map shows the cavity mode and its dispersion characteristic. (b) Cavity dispersion in the $x$-direction (red) or $y$-direction (blue), for the example HCG-based cavity. The calculated dispersion curvatures, $\partial^2\omega/\partial k_{x,y}^2$ are 64.2 and 266.9 m$^2$/s for the $x$- and $y$-directions, respectively.}
\label{fig:Dispersion}
\end{figure}

Figure \ref{fig:Dispersion}(a) illustrates the transmissivity contour map of the test HCG-based cavity as a function of the wavelength and incident angle in the $x$-direction. As explained in section \ref{sec:QFactorMethods}, the mode resonance frequency $\omega$ corresponds to the peak in the transmissivity spectrum, which is shown as the bright line in the contour map. Since the dispersion calculation requires the cavity mode frequency graph as a function of in-plane wavevector, all three methods explained in section \ref{sec:QFactorMethods} can be applied. The dispersion graphs computed with QNM approach for the $x$- and $y$-directions are illustrated in Fig. \ref{fig:Dispersion}. The Fabry-Perot and spectral methods will also lead to the same dispersion graphs. The dispersion curvatures for $x$- and $y$-directions, i.e. $\partial^2\omega/\partial k_{x,y}^2$, are calculated to be 64.2 m$^2$/s and 266.9 m$^2$/s, respectively. 

\subsection{Low-Dimensional Simulation Examples}
In order to prove the accuracy of low-dimensional simulations in modeling the HCG-based cavity, we perform 2D, 2.5D and 3D simulations of the test structure. 
Table \ref{tab:LowerDSimulations} shows the simulation results. By performing a 2D and a 1.5D simulation of the structure to estimate $Q_{2D}$ and $Q_z$, respectively, $Q_x$ can be determined: $1/Q_{2D}=1/Q_z+1/Q_x$. Similarly, a 2.5D simulation estimates $Q_{2.5D}$, which determines $Q_y$ by using the relation, $1/Q_{2.5D}=1/Q_z+1/Q_y$. Then, by inserting the $Q_x$, $Q_y$, and $Q_z$ values into Eq. (\ref{eq:DirectionQFactor}b), the total Q-factor $Q$ can be estimated. As shown in Table \ref{tab:LowerDSimulations}, this $Q$ value determined from low-dimensional simulations agrees well with the $Q_{3D}$ value that is directly obtained from a 3D simulation. The relative difference between these two Q-factors is less than 1\%. In addition, once the band-edge frequency and band-edge curvature are known for the core and cladding of the vertical cavity, the original 3D problem becomes a 2D photonic well problem, from which the resonance wavelengths of the modes can be estimated as explained in Ref. \cite{Taghizadeh2015a}. Therefore, the resonance wavelengths and Q-factor of modes can be estimated without performing a 3D simulation. The time required for performing all 1.5D, 2D and 2.5D simulations is approximately two orders of magnitude shorter than that for a single 3D simulation. 

\begin{table}
\renewcommand{\arraystretch}{1.3}
\caption{Low-dimensional simulation results for the example HCG-based vertical cavity. }
\label{tab:LowerDSimulations}
\centering
\begin{tabular}{c | c | c | c} 
\hline
Simulation & $\lambda_r$ (nm) & Q-factor & Result in \\ 
\hline
1.5D & 1550.00 & $Q_{1.5D}$=118098 & $Q_z$=118098 \\ 
2D & 1549.0621 & $Q_{2D}$=4777 & $Q_x$=4978 \\
2.5D & 1546.1032 & $Q_{2.5D}$=4622 & $Q_y$=4811 \\
3D & 1545.26 & $Q_{3D}$=2308 & Eq. [\ref{eq:DirectionQFactor}(b)] $Q$=2303 \\ \hline
\end{tabular}
\end{table}

\subsection{In-plane Heterostructure}
\begin{figure}[t]
\centering
\includegraphics[width=0.43\textwidth]{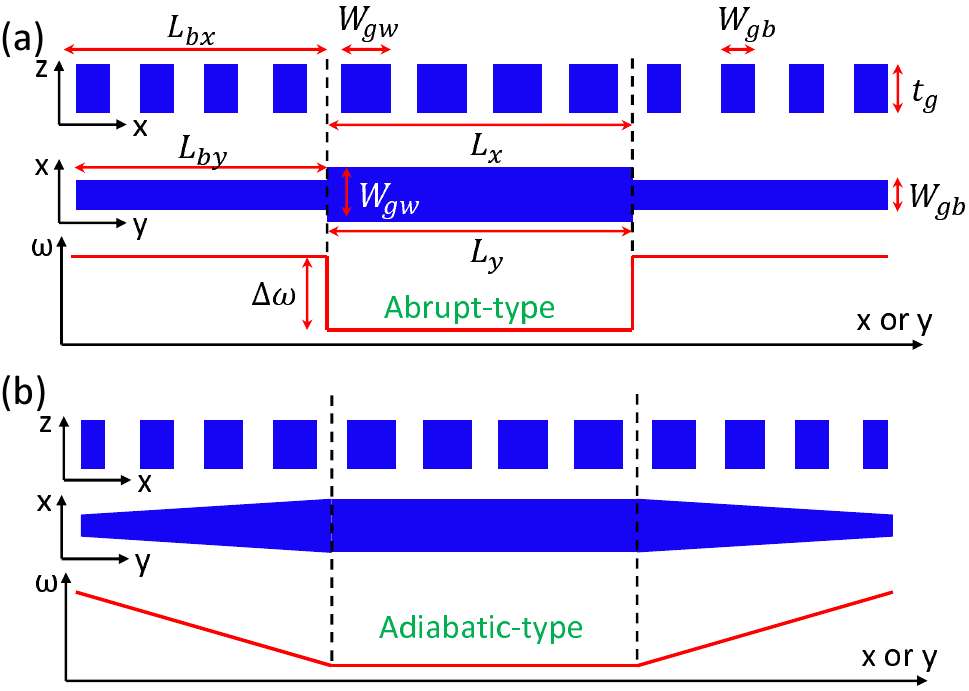}
\caption{(a) Cross-sectional schematic of a grating bar in the $x$-$z$ plane (top), and $x$-$y$ plane (middle) for HCG-based vertical cavity with an in-plane heterostructure, which forms as a photonic well with (a) Abrupt-type barrier, and (b) Adiabatic-type barrier. The grating bar width $W_g$ is varied along the $x$- and $y$-directions to make a heterostructure with the well sizes of $L_x$ and $L_y$. The corresponding spatial profile of energy bandedge for the heterostructure is shown schematically below it.}
\label{fig:HeteroSch}
\end{figure}

\begin{figure}[t]
\centering
\includegraphics[width=0.35\textwidth]{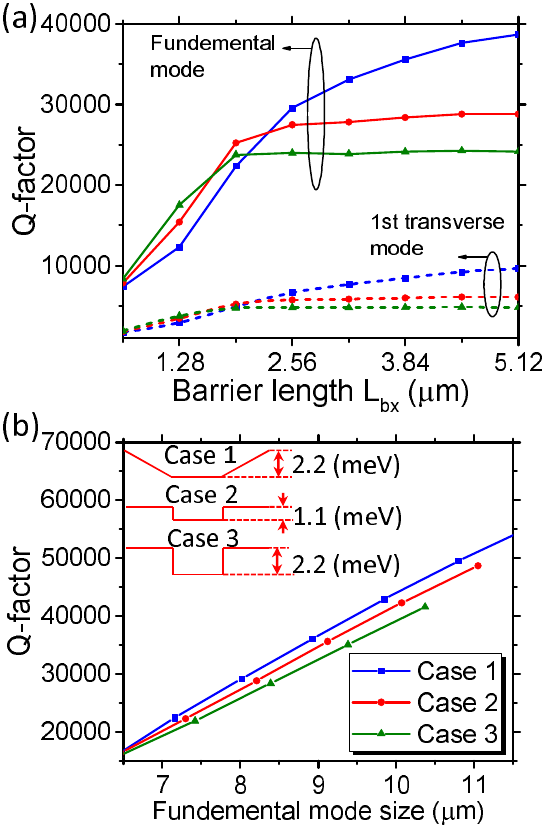}
\caption{(a) Q-factors of the fundamental and first-order transverse mode in the example HCG-based cavity structure as a function of barrier length $L_{bx}$, for a abrupt-barrier with height of 1.1 meV (in blue), 2.2 meV (in red), and 3.3 meV (in green). (b) Q-factors of the fundamental mode in (a) versus its mode size for several different barrier cases. The mode size is defined as the width, where the field amplitude drops to $1/e$ of its maximum value. } 
\label{fig:AbruptAdiabatic}
\end{figure}

An in-plane heterostructure can be formed in a HCG-based cavity by introducing variations of the grating parameters in the $x$- and $y$-directions, as shown in Fig. \ref{fig:HeteroSch} \cite{Taghizadeh2015a}. These variations can be realized by changing the grating bar width in the e-beam lithography process. Here, we are particularly interested in photonic well structures for implementing efficient transverse mode confinement, which is essential for laser applications. In this section, it is discussed how to design an effective barrier in the $x$-direction, with a similar approach being applicable for the $y$-direction. In a vertical cavity heterostructure, there are three different loss mechanisms; vertical out-coupling loss through mirrors, lateral out-coupling loss through the in-plane barriers, and scattering loss due to the refractive-index perturbation of the cavity by the heterostructure \cite{Sciancalepore2011}. For our example 2D HCG-based cavity with a heterostructure in the $x$-direction only, both mirrors are designed to be highly-reflective (reflectivity above 99.99\%), which results in a low vertical out-coupling loss. Figure \ref{fig:AbruptAdiabatic}(a) shows the Q-factors of two transverse modes as a function of the barrier length $L_{bx}$ for three barrier heights $\Delta\omega$. In all cases, the Q-factor saturates as the barrier length $L_{bx}$ increases. A smaller barrier height requires a longer barrier length for saturation, and results in a higher Q-factor since the scattering loss becomes smaller. The barrier length in the transverse directions should be at least several times the barrier penetration depth, in order to have an effective barrier. Otherwise, the optical mode tail will tunnel through the barrier region and leak out, resulting in additional lateral loss. For a large-enough barrier length, a smaller barrier height is advantageous, since it results in a smaller perturbation and consequently less scattering loss. However, it should be noted that the optical mode size will be larger for a smaller barrier height, since it will penetrate farther into the barrier region. 

The barriers can be formed adiabatically or abruptly, as shown in Figs. \ref{fig:HeteroSch}(a) and (b), respectively. 
Figure \ref{fig:AbruptAdiabatic}(b) compares the Q-factor of the fundamental mode as a function of mode size, barrier height, and barrier type (abrupt or adiabatic), given the barrier lengths are identical. Comparing Cases 1 and 3 shows that the adiabatic barrier leads to a higher Q-factor and a larger mode size. The higher Q-factor can be attributed to less scattering loss. Comparison of Cases 2 and 3 shows that a smaller barrier height results less scattering loss and a larger mode size.  


\section{Conclusion}
\noindent 
The HCG-based vertical cavity laser is an emerging micro-cavity structure with promising applications. However, the modeling of its optical characteristics is highly demanding, due to the lack of cylindrical symmetry in the structure, rendering it a genuine 3D problem with vectorial nature. In this work, to tackle this numerical challenge, a fully-vectorial 3D simulator, based on the Fourier modal method, has been implemented and optimized using structural symmetries and absorbing boundary conditions. Using this simulator, several techniques for computing the Q-factor and resonance frequency have been compared, which shows that the Fabry-Perot approach is the most efficient for handling HCG-based cavities with a few transverse modes. It also reveals that the uncertainty in the Q-factor calculations is inherently larger by a factor given by the Q-factor itself compared to the uncertainty of the resonance frequency calculation. Furthermore, a method for analyzing a 3D structure by performing consecutive lower-dimensional simulations, has been proposed and tested. This method can accurately estimate the Q-factor, with much less computational effort, i.e. simulation time and memory, compared to a 3D simulation. Finally, using the numerical techniques discussed here, various in-plane heterostructures for transverse confinement have been investigated, which shows that adiabatically-defined heterostructure reduces the optical loss as well as providing good transvese confinement.


\section*{Acknowledgment}
The authors gratefully acknowledge support from the Danish Council for Independent Research (Grant No. 0602-01885B), the Innovation Fund Denmark through the HOT project (Grant No. 5106-00013B), as well as Villum Fonden via the NATEC Centre of Excellence.

\ifCLASSOPTIONcaptionsoff
  \newpage
\fi



\def\url#1{}
\bibliographystyle{IEEEtran}
\bibliography{Simulator}

\begin{IEEEbiography}[\addportrait{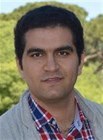}]{Alireza Taghizadeh}
received the two simultaneous B.S. degrees in both Electrical Engineering and Physics and the M.S. degree in Electrical Engineering, from the Isfahan University of Technology, Isfahan, Iran in 2009 and 2011, respectively, and the Ph.D. degree in Photonics Engineering from the Technical University of Denmark (DTU Fotnik), Lyngby, Denmark, in 2016. Since 2016, he has been a Post-Doctoral researcher at DTU Fotnik. His current research interests include computational photonics, semiconductor device physics, and in particular design, modeling and numerical simulation of novel micro/nano lasers.
\end{IEEEbiography}

\begin{IEEEbiography}[\addportrait{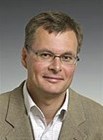}]{Jesper M{\o}rk}
received the M.S., Ph.D., and Dr. Techn. degrees from the Technical University of
Denmark (DTU Fotnik), Lyngby, Denmark, in 1986, 1988, and 2003, respectively.
Since 2002, he has been a Professor of Semiconductor Photonics, and since 2008 he is heading the nanophotonics section at the Department of Photonics Engineering, DTU Fotonik, formerly known as Research Center COM. He is the author of more than 220 papers in refereed journals and 300 contributions to international conferences, including 80 invited talks. His current research interests include semiconductor device physics, in particular ultrafast devices for optical signal processing, nanolasers, other photonic devices and quantum photonics.
\end{IEEEbiography}


\begin{IEEEbiography}[\addportrait{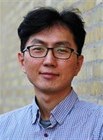}]{Il-Sug Chung}
received the B.S. and M.S degrees in physics from the Korea Advanced Institute of
Science and Technology, Korea, in 1997 and 2000, respectively, and the Ph.D. degree in optoelectronics from the Gwangju Institute of Science and Technology, Korea, in 2006. Since 2006, he has been with the Department of Photonics Engineering at Technical University of Denmark, Denmark, and is currently Associate Professor. His current research interests cover the modeling, fabrication, and characterization of micro/nano lasers for Si photonics, ultrahigh-speed optical communications, and sensing applications.
\end{IEEEbiography}


\vfill


\end{document}